\documentclass{aa}  

\usepackage{graphicx}
\begin{document}

   \title{The signature of primordial black holes in the dark matter
          halos of galaxies}


   \author{M.R.S. Hawkins
          \inst{1}
          }

   \institute{Institute for Astronomy (IfA), University of Edinburgh,
              Royal Observatory, Blackford Hill,
              Edinburgh EH9 3HJ, UK\\
              \email{mrsh@roe.ac.uk}
             }

   \date{Received 5 August, 2019 / Accepted 17 October 2019}

 
  \abstract
   {}
   {The aim of this paper is to investigate the claim that stars in the
    lensing galaxy of a gravitationally lensed quasar system can always
    account for the observed microlensing of the individual quasar
    images.}
   {A small sample of gravitationally lensed quasar systems was chosen
    where the quasar images appear to lie on the fringe of the stellar
    distribution of the lensing galaxy.  As with most quasar systems, all
    the individual quasar images were observed to be microlensed.  The
    surface brightness of the lensing galaxy at the positions of the
    quasar images was measured from {\it Hubble} Space Telescope frames, and
    converted to stellar surface mass density.  The surface density of
    smoothly distributed dark matter at the image positions was obtained
    from lensing models of the quasar systems and applied to the stellar
    surface mass density to give the optical depth to microlensing.  This
    was then used to assess the probability that the stars in the lensing
    galaxy could be responsible for the observed microlensing.  The
    results were supported by microlensing simulations of the star fields
    around the quasar images combined with values of convergence and shear
    from the lensing models.}
   {Taken together, the probability that all the observed microlensing is
    due to stars was found to be $\sim 3 \times 10^{-4}$.  Errors resulting
    from surface brightness measurement, mass-to-light ratio, and the
    contribution of the dark matter halo do not significantly affect this
    result.}
   {It is argued that the most plausible candidates for the microlenses
    are primordial black holes, either in the dark matter halos of the
    lensing galaxies, or more generally distributed along the lines of
    sight to the quasars.}

   \keywords{dark matter -- gravitational lensing: micro -- galaxies: halos}

   \maketitle
%

\section{Introduction}

The identification of the material making up the dark matter component
of the Universe remains one of the most significant unsolved problems
in cosmology.  So far, most of the effort has gone into the search for
elementary particle candidates, but with little success to date.  The
limits on dark matter particles come from a wide variety of sources,
including direct detection \citep{a12,a14b}, annihilation signals
\citep{a14a} and cosmic microwave background (CMB) measurements
\citep{m14}.  The search for dark matter in other forms has received
little attention, and has largely been confined to the Galactic halo
\citep{a00,t07,w11}, with inconclusive and apparently contradictory
results \citep{h15}.  In these searches, the idea is to detect a
population of compact bodies in the halo of the Milky Way by observing the
rate at which stars in the background Magellanic Clouds are microlensed.
The detection of gravitational waves from a black hole merger \citep{a16}
has again brought into prominence the possibility \citep{h93,h11} that
dark matter may be in the form of compact bodies, which would most
plausibly be solar mass primordial black holes \citep{b16,s16}.

\begin{figure}
\begin{picture}(200,200)(-5,-10)
\includegraphics[width=0.46\textwidth]{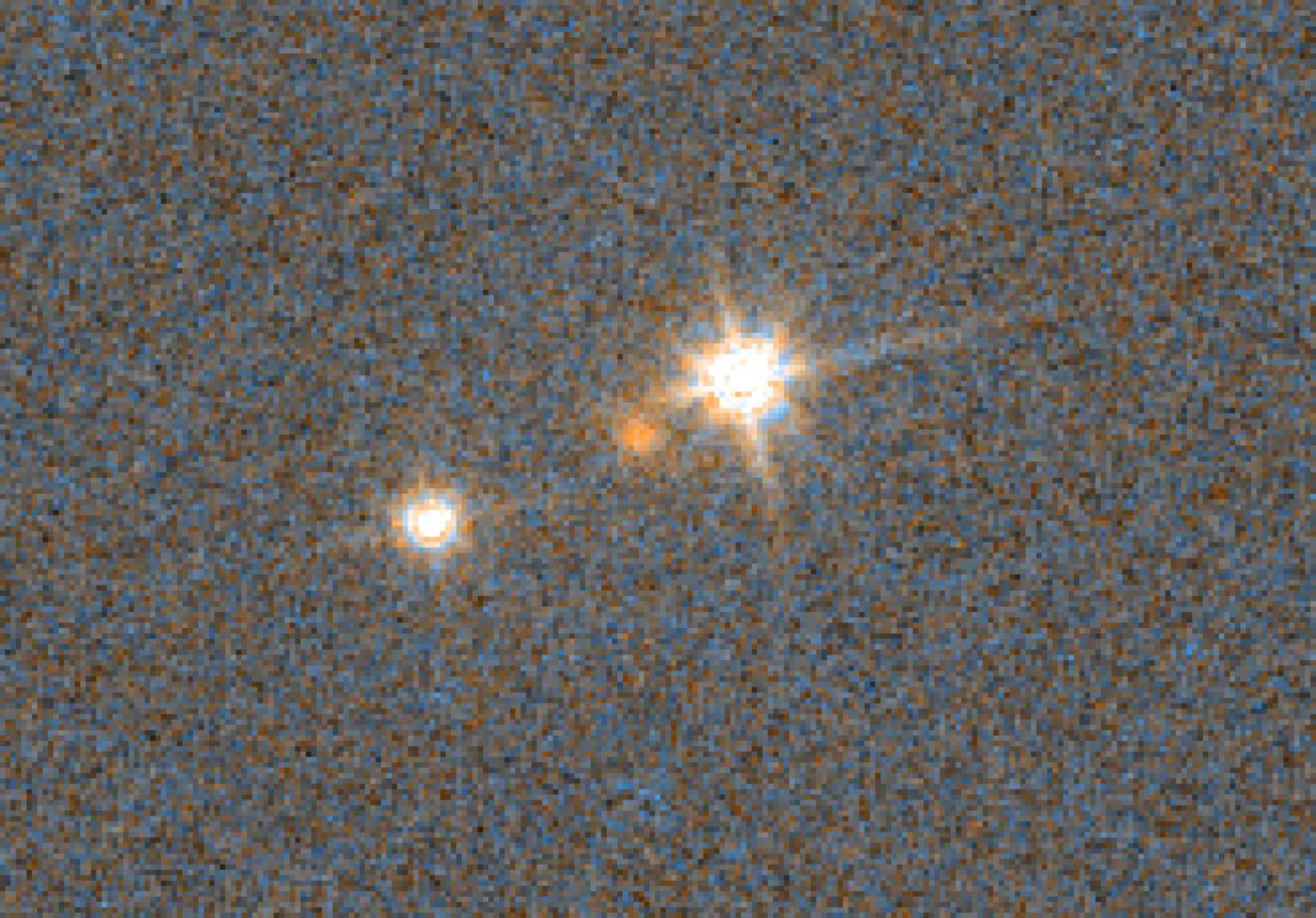}
\end{picture}
\caption{{\it Hubble} Space Telescope image in F555W band of lensed
 quasar HE1104-1805 showing lensing galaxy.}
\label{fig1}
\end{figure}

One of the most interesting predictions of the theory of general
relativity is the creation of multiple images by gravitational lensing
\citep{e36}.  A massive galaxy close to the line of sight to a quasar
typically splits the image into  two or four observable components.  This
phenomenon was first observed by \cite{w79} where the image of the quasar
Q0957+561 was split into two components by a massive foreground galaxy.
This led to the interesting prediction \citep{c79} that stars along the
light paths to each image could cause significant flux changes on a
timescale of a few years.  \cite{g81} used this microlensing effect to
propose a method to determine whether the dark halos of galaxies were
composed of compact bodies.  Microlensing can be distinguished from
intrinsic variations of the quasar where the changes in brightness in each
image are the same, separated by a short time interval to reflect the
difference in light travel time to the observer.  For microlensing
variations, the individual images change in brightness independently
of each other, typically on a timescale of a few years.  This is due to
the different trajectories of the light rays to each image that are
microlensed by different magnification patterns corresponding to the
distributions of compact bodies through which they have passed.
Microlensing was first definitively observed by \cite{s91}, where the two
images of Q0957+561 were seen to vary independently of each other.  Since
then microlensing has been observed in most quasar systems which have been
photometrically monitored over a period of a few years.  At first sight
this would appear to be a powerful argument for a cosmological
distribution of compact bodies acting as microlenses.  Given that every
line of sight appears to be microlensed, this population of compact bodies
would be sufficient to make up the dark matter component of the Universe
\citep{p73}.  There is however a problem with this conclusion.  Lensed
quasars from their nature must have a massive galaxy along the line of
sight, and it has generally been assumed that the stellar population in
these galaxies is responsible for the observed microlensing
\citep{s90,f91,k93}.  However, casual inspection of some well-studied
lensed quasar systems seems to cast doubt on this hypothesis, the images
appearing to lie well clear of the main light concentration of the lensing
galaxy.  A good example of this is the double image quasar system
HE1104-1805, shown in Fig.~\ref{fig1}.

The question of the proportion of dark matter in lenses for multiple
quasar systems has been addressed by a number of groups
(e.g. \cite{m09,p12}).  The approach used in these papers is to evaluate
the likelihood of microlensing measurements using frequency distributions
obtained from simulated microlensing magnification maps for different
values of the fraction of mass in compact objects.  They find that maximum
likelihood tests favour a mass fraction of less than 10\%, which is
consistent with the stellar populations of the lensing galaxies.  The
samples of quasars used in these surveys are quite large, and mostly
comprise systems where the quasar images are deeply embedded in the
lensing galaxies, and where there is no reason to doubt that the stellar
population can account for the observed microlensing.

In this paper we approach the problem of identifying the objects
responsible for the microlensing events by starting with observations of
the stellar population of the lensing galaxy.  We investigate whether the
optical depth to microlensing $\tau$ associated with these stars is
sufficient to produce the observed microlensing.  There are several
well-studied quasar systems for which the stellar population of the
lensing galaxy provides a plausible source for the microlenses. In the
Einstein Cross quadruple system Q2237+0305 which has been extensively used
for microlensing studies the four images lie close to the centre of the
lensing galaxy, and there seems to be little doubt that the stars can be
responsible for the observed microlensing.  Similarly, image B of
Q0957+561 lies deep inside the lensing galaxy, and again the microlensing
could plausibly be produced by the stars.  There are however a number of
systems where the quasar images are being microlensed, but due to the
alignment and redshifts of the source and lens, and the mass and
compactness of the lensing galaxy, the quasar images lie on the fringe of
the stellar distribution.  In such cases it is not clear that the optical
depth to microlensing of the stars is sufficient to produce the observed
microlensing.  Complicating factors in the calculation of $\tau$ include
the possible presence of caustic-induced features that start to
appear at moderate optical depth \citep{k97,l97}, and the effect of any
smoothly distributed dark matter in the halo, which increases the
effective optical depth to microlensing due to stars or any other compact
bodies \citep{w92,s02}.

This study is based on four gravitational lens systems where the image of
a quasar is split into two or four components by a massive galaxy, and the
images appear to lie on the fringe of the stellar distribution.  All
four systems have well-sampled light curves which show large amplitude
uncorrelated variations between the quasar images that are attributed
to microlensing.  The procedure we adopt is to measure the surface
brightness of the lensing galaxies as a function of galactocentric
distance from HST frames, and convert this to stellar surface mass
density.  To obtain the total mass profile, and hence the dark matter
distribution, we use lens models based on the software of \cite{k01}.
This also gives values for convergence and shear at the image positions.
Assuming the dark matter to be smoothly distributed, this can then be used
to convert stellar surface brightness to optical depth to microlensing.
To assess the effects of shear on the microlensing probability we generate
simulated microlensing magnification maps.  These simulations also provide
a way of checking the validity of microlensing probabilities derived from
the optical depth measures.  An additional correction is needed to account
for the finite size of the quasar source.  For this purpose we use the
most recent values from the literature.

The results of this analysis imply that the probability of the observed
microlensing in any one quasar system being due to the stellar population
of the lensing galaxy is quite small.  We are however mainly interested in
the probability that the microlensing in at least one system is not due
to stars.  We thus ask what is the probability that the microlensing in
all the quasar systems in our sample can be attributed to stars, and find
the probability to be very small, of the order of $3 \times 10^{-4}$.  For
the lensing galaxies to be the source of the microlensing objects, a large
part of the dark matter halo must be in the form of stellar mass compact
bodies.  We conclude that either the dark matter content of the galaxy
halos is made up of non-stellar compact bodies, or that there is a more
general near-critical density distribution of compact bodies along the
line of sight to the quasar, which would make up the dark matter content
of the Universe.

If stars in the lensing galaxy are not responsible for the observed
microlensing and the variations are attributed to a cosmological
distribution of lenses, this raises the question of what the compact
bodies really are. If they make up the dark matter then baryonic
candidates, including stars and their evolutionary products such as white
dwarfs, are ruled out on the basis of baryon synthesis constraints
\citep{s93}.  A number of non-baryonic candidates have been proposed
during the last 30 years or so but we argue that among these, primordial
black holes are the only objects for which a plausible case can be made
that they make up the dark matter.  Throughout the paper we assume
$\Omega_M = 0.27$, $\Omega_\Lambda = 0.73$ and
$H_0 = 71$ km s$^{-1}$ Mpc$^{-1}$ for cosmological calculations.

\section{The quasar sample}
\label{sec2}

\begin{table*}
\caption{Basic data for lensed quasar systems.}
\label{tab1}
\begin{center}
\vspace{5mm}
\begin{tabular}{lcccccccccc}
\hline\hline
 & & & & & & & & & & \\
 System & $z_q$ & $z_g$ & $m_I$ & $M_I$ & $F_1/F_2$ & $\Delta_0$ &
 $\Delta m$ & {$R_c$} & $R_E$ & Ref. \\
 & & & & & & & & (kpc) & ($10^{16}$ cm) & \\
\hline
 & & & & & & & & & & \\
 HE1104-1805  & 2.32 & 0.73 & 20.01 & -22.65 & A/B & -1.13 & 0.57 &
 6.81 & 1.97 & 1 \\
 HE0435-1223  & 1.69 & 0.46 & 18.05 & -23.57 & A/B & -0.38 & 0.41 &
 5.05 & 2.44 & 2 \\
 RXJ1131-1231 & 0.66 & 0.29 & 17.88 & -22.69 & A/C & -1.12 & 1.02 &
 4.37 & 2.09 & 3 \\
 WFI2033-4723 & 1.66 & 0.66 & 19.71 & -22.72 & B/C & -0.10 & 0.49 &
 5.72 & 1.97 & 4 \\
 & & & & & & & & & & \\
\hline
\end{tabular}
\end {center}
{\bf References.} (1) \cite{o03,p07}: (2) \cite{c11}; (3) \cite{t13c};
 (4) \cite{v08,t13a}
\end{table*}

\begin{figure*}
\centering
\begin{picture} (0,400) (260,0)
\includegraphics[width=1.0\textwidth]{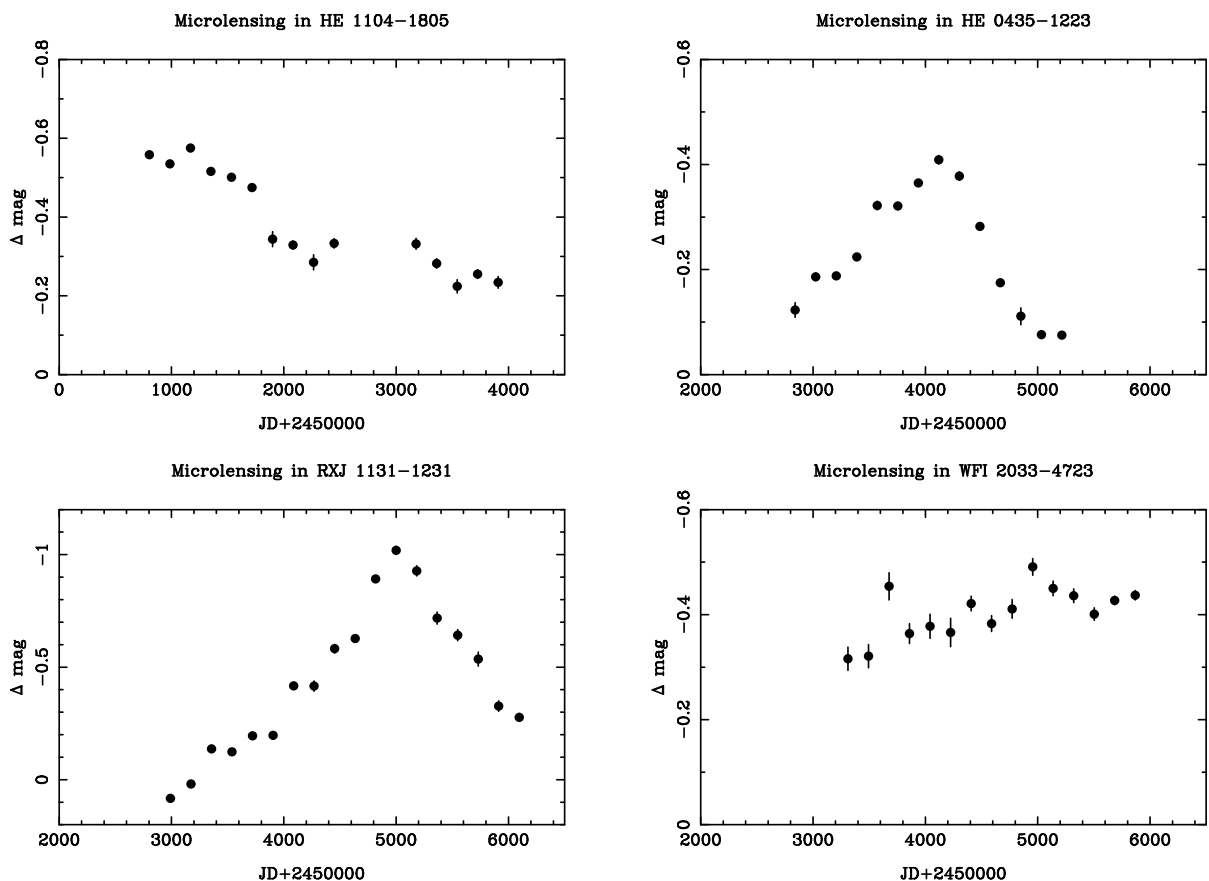}
\end{picture}
\caption{Difference light curves for images of gravitationally
 lensed quasars.  The differential variations of the images
 are attributed to microlensing.  The derivation of the zeropoints is
 described in the text.}
\label{fig2}
\end{figure*}

The criteria for the selection of the sample of quasar systems used
in this study were designed to choose systems where the images lie on the
outskirts of the observable part of the lensing galaxy.  In
gravitationally lensed quasar systems, several factors determine the
position of the images relative to the stellar distribution in the galaxy.
The separation of the images on the sky is proportional to the deflecting
mass, and their angular distance from the galaxy centre depends on the
true angular separation of the source and lens, and their redshifts
\citep{s92}.  The compactness of the lensing galaxy also clearly
affects the extent to which the quasar images lie within the stellar
distribution.  The parent sample for the selection of quasar systems
to be used in this study was the
CASTLES\footnote{https://www.cfa.harvard.edu/castles/} database.  The
criteria for inclusion were that the size estimate as given by CASTLES
should be greater than 2.0 arcsec, and that the $I$ magnitude of the
lensing galaxy should be less than 20.0.  The first of these criteria was
to maximise the likelihood that the quasar images lie clear of the stellar
distribution of the lensing galaxy, and the second to enable accurate
photometry of the stellar population.  This resulted in a sample of six
quasar sytems.  The second criterion was that there should be well sampled
light curves available, typically covering around seven years or more.  Of
the preliminary sample five comfortably passed this test, but the
available photometric data for PG1115+080 were fragmentary and inadequate
for our programme.  Of the remaining five systems, although Q0957+561
satisfied the criteria to be included in the sample, image B is deeply
embedded in the lensing galaxy with a measured optical depth to
microlensing $\tau \sim 0.6$ from the stars alone.  For this system there
is no reason to believe that the stars are not responsible for the
microlensing and so it was not included in the analysis.  The final sample
of four systems is shown in Table~\ref{tab1}, together with some basic
parameters.  $z_q$ and $z_d$ are the redshifts of the lensed quasar and
the lensing galaxy respectively, $m_I$ is the apparent magnitude of the
lensing galaxy in the $I$-band from CASTLES measurements, and $M_I$ the
associated absolute magnitude.  Also shown in Table~\ref{tab1} is the core
radius $R_c$ containing half the observed light and the Einstein radius
$R_E$ for a $0.2 M_{\odot}$ compact body.

Evidence for microlensing has been found in most gravitationally lensed
quasar systems which have been studied in detail.  There are basically
two ways in which microlensing has been detected.  Monitoring of lensed
quasar systems shows that the individual images usually vary in an
uncorrelated way, which is generally accepted to be the result of
microlensing by compact bodies along the line of sight.  Much effort has
been been put into monitoring brightness changes, notably by the
COSMOGRAIL collaboration.  This project was primarily motivated by the
challenge of measuring time delays between intrinsic brightness changes
in the quasar images of a lensed system for the purpose of estimating the
value of the Hubble constant \citep{e05}.  However, the light curves
produced by the group form an excellent database for measuring the effects
of microlensing \citep{t13b}.  An alternative approach is to compare the
luminosity of the accretion disc which is small enough to be microlensed
by solar mass bodies with flux from less compact regions where little or
no microlensing would be expected.  A good example is the broad line
region which is generally taken to be too large for microlensing by
stellar mass objects to produce significant changes in brightness. The
ratio of broad line fluxes in the quasar images is used to provide a datum
for variations in continuum flux, enabling a single epoch measurement of
microlensing \citep{m12,s12}.  An alternative approach is to use flux
ratios of the quasar images in the infrared on the assumption that they
originate from an area too large to be microlensed \citep{b11,s12}. One
of the shortcomings of measuring microlensing from differential flux
in the component images is that it does not provide an absolute
measurement of the resulting change in flux ratio.  This problem can be
largely overcome by using the single epoch measures of microlensing as
described above to provide a datum for the non-microlensed image
brightness ratios \citep{s12}.

All of the quasar systems in our sample in Table~\ref{tab1} have been
the subject of extensive optical monitoring programmes for the purpose of
measuring time delays in light travel time between the images.  For
HE1104-1805, early work by \cite{g02} showed clear evidence for
differential variation between the two images over a period of around 3
years.  This was followed by longer runs of closely sampled data
\citep{s03,o03,p07} to give photometry covering some 8 years.  The first
light curves for HE0435-1223 covered 2 years \citep{k06} and already
showed clear signs of microlensing in the differential variation of the
images.  This work was followed by a large monitoring programme undertaken
by the COSMOGRAIL collaboration spanning 7 years \citep{c11} which we use
here.  RXJ1131-1231 is another target monitored by COSMOGRAIL
\citep{t13c}, and shows the largest microlensing amplitude in our sample,
of more than a magnitude.  The final member of our sample, WFI2033-4723,
was monitored by COSMOGRAIL for three years after which they found no
evidence for microlensing \citep{v08}.  However, after extending the
observations to 8 years \citep{t13a} a variation of around 0.2 magnitudes
was detected.  The presence of microlensing was confirmed by differences
between the flux ratios of broad band and emission line measures, which
implied a microlensing signal of $\sim 0.5$ magnitudes.
 
Fig.~\ref{fig2} shows difference light curves for images of the lensed
quasar systems listed in Table~\ref{tab1}.  The references below the table
are for the monitoring data on which the difference curves are based.  For
the purpose of uniform comparison, and to highlight the long timescale of
the microlensing variations, the original data are plotted as half-yearly
weighted means.  The image pairs corresponding to the difference curves in
Fig.~\ref{fig2} (as named in the references to the light curves) are given
as $F_1/F_2$ in Table~\ref{tab1}, and the zeropoints for the flux ratios
in the absence of microlensing are shown as magnitude difference
$\Delta_0$.  These zeropoints are derived from the flux ratios of the
broad emission line region or from infrared photometry where the emission
region is supposed to be too large to be significantly microlensed by
stellar mass objects.  The zeropoint for HE1104-1805 is taken from
\cite{m12}, and for the remainder from \cite{s12} (spectroscopic data) and
\cite{b11} (infrared photometry).  In all cases zeropoints from the two
methods are in good agreement.  All four systems show distinctive
microlensing variations on a variety of timescales, consistent with the
statistics of recent surveys for microlensing \citep{m12,s12}.  The
lightcurves of HE1104-1805 and WFI2033-4723 are dominated by long
timescale variations of at least 20 years, while HE0435-1223 and
RXJ1131-1231 contain complete events with a timescale of around 10 years.
Adopting the conventionally assumed mean transverse velocity of
600 km sec$^{-1}$ for the microlenses, this corresponds to lens masses of
around 0.3 M$_{\odot}$.  The amplitude $\Delta m$ of the light curves in
magnitudes is also given in Table~\ref{tab1}.

\section{The lensing galaxies}
\label{sec3}

\begin{figure}
\centering
\begin{picture} (0,200) (130,0)
\includegraphics[width=0.5\textwidth]{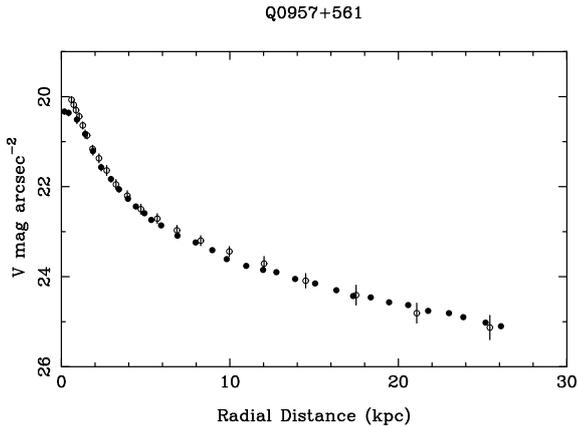}
\end{picture}
\caption{Surface brightness profile for the lensing galaxy of Q0957+561
 from a WFPC2 image in the F555W band.  Open circles show measures
 obtained using point spread function subtraction of the quasar
 images from \protect\cite{b97}, and filled circles are from the present
 work.}
\label{fig3}
\end{figure}

\begin{figure*}
\centering
\begin{picture} (0,200) (260,0)
\includegraphics[width=1.0\textwidth]{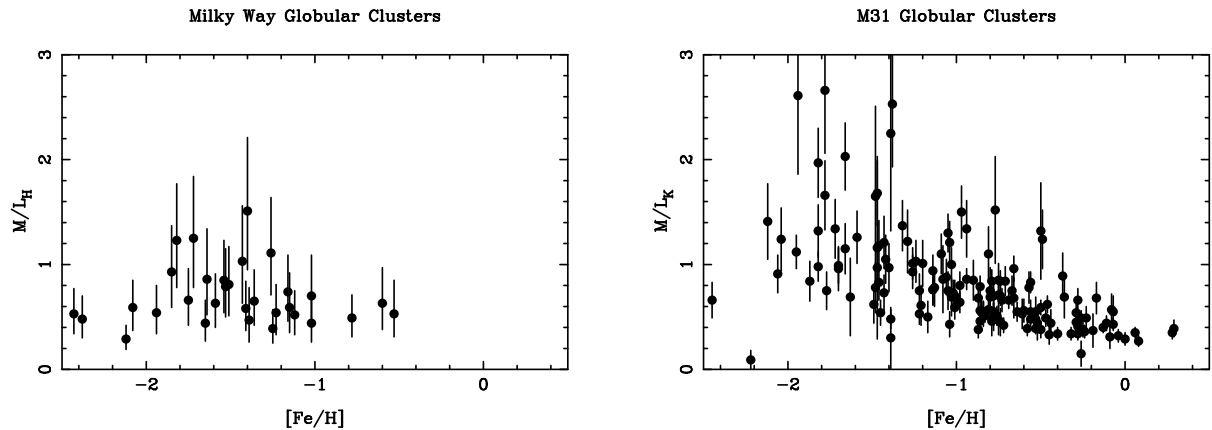}
\end{picture}
\caption{$M/L_H$ versus [Fe/H] for Milky Way globular clusters based on
 measures of $M/L_V$ from \protect\cite{m05a} and photometry from
 \protect\cite{c07} (left hand panel), and $M/L_K$ versus [Fe/H] for M31
 globular clusters from \protect\cite{s11} (right hand panel).}
\label{fig4}
\end{figure*}

\begin{figure*}
\centering
\begin{picture} (0,400) (260,0)
\includegraphics[width=1.0\textwidth]{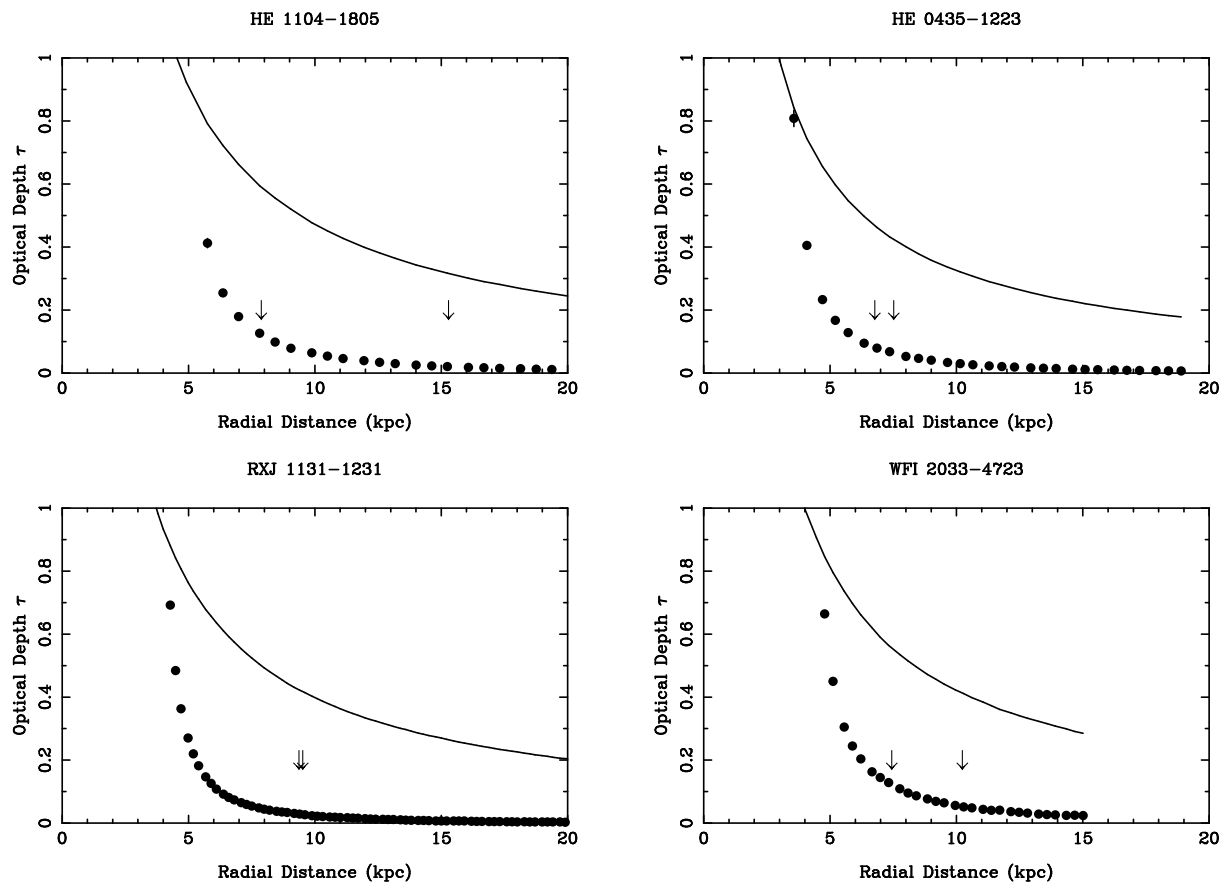}
\end{picture}
\caption{Optical depth to microlensing $\tau$ as a function of radial
 distance for the lensing galaxies of four gravitationally lensed
 quasars.  The solid line shows the surface mass density in smoothly
 distributed matter as discussed in the text, and the filled circles show
 the optical depth to microlensing for the stellar population of the
 lensing galaxy.  The arrows indicate the positions of the quasar images
 listed in Table~\ref{tab2}.}
\label{fig5}
\end{figure*}

In order to test the idea that the stellar population of the lensing
galaxy is responsible for the observed microlensing, we built up a mass
distribution for the lensing galaxy by sticking as far as possible to
observations.  We started by converting galaxy surface brightness
measurements to stellar surface mass density, and hence to optical depth
to microlensing $\tau$.  For large values of $\tau$ a complex network of
caustics is formed, resulting in a non-linear amplification pattern for
the source.  However, for low values, $\tau$ is essentially the
probability that the trajectory of a light ray is significantly
amplified by the gravitational lensing effect of a compact body along the
line of sight. A criterion for this low optical depth regime is that
$\tau \lesssim 0.1$ \citep{k97,l97}.  $\tau$ depends on the surface mass
density of both the compact lensing bodies and the more smoothly
distributed matter \citep{s92}, and is defined by:

\begin{equation}
\tau = \frac{\kappa_*}{1-\kappa_c}
\label{eqn1}
\end{equation}

\noindent
where $\kappa_*$ and $\kappa_c$ are the surface mass densities in compact
bodies and smoothly distributed matter respectively, in units of the
critical surface mass density $\Sigma_{cr}$, and

\begin{equation}
\Sigma_{cr} = \frac{c^2}{4 \pi G} \frac{D_q}{D_g D_{gq}}
\label{eqn2}
\end{equation}

\noindent
$D_g$, $D_q$, and $D_{gq}$ are the angular diameter distances to the
lensing galaxy, to the quasar, and from the galaxy to the quasar
respectively. To calculate $\tau$ we need values for $\kappa_*$ and
$\kappa_c$.

\subsection{Surface brightness}

To evaluate $\kappa_*$, we first measured the surface brightness profiles
of the lensing galaxies.  For this purpose we used HST frames taken with
the F160W filter, close to the $H$-band, for the four systems in
Table~\ref{tab1}.  Working in the infrared, as opposed to the optical,
provides a much more favourable contrast between the quasar images which
are typically blue, and the redshifted lensing galaxies which are bright
in the infrared.  Standard IRAF routines were used to measure the F160W
flux as a function of distance from the galaxy centre.  To recover the
undistorted profile of the lensing galaxy, IRAF image analysis routines
were used to subtract the flux from the contaminating quasar images and
derive the underlying galaxy surface brightness.  Most lensing galaxies
are symmetric across diagonals, and this feature was used to evaluate the
flux contribution from the quasar images to be subtracted from the
contaminated galaxy image, and interpolate across spurious small features
to provide a clean galaxy profile.  These flux measurements were then
converted to surface brightness using the appropriate HST photometric
zeropoints.  In order to check the accuracy of this procedure, the results
were compared with similar work from the literature.  No surface
brightness profiles for the four systems in Table~\ref{tab1} have so far
been published, but \cite{b97} have measured the profile of the heavily
contaminated galaxy in the well-known binary system Q0957+561.  This
system has not been used for the present study because image B lies close
to the centre of the lensing galaxy, and could plausibly be microlensed by
the stellar population.  However it makes a good test bed for image
subtraction techniques.  Fig.~\ref{fig3} shows the surface brightness
profile of the lensing galaxy of Q0957+561 from an HST frame as measured
by \cite{b97}, together with measurements of the same frame using the
methods employed for this paper.  The agreement is satisfactory, and any
discrepancies are small compared with other uncertainties in the
evaluation of $\tau$.

\subsection{Mass-to-light ratio}

\begin{table*}
\caption{Microlensing optical depth for galaxy halos.}
\label{tab2}
\centering
\vspace{5mm}
\begin{tabular}{lcrccccc}
\hline\hline
 & & & & & & & \\
 System & Image & \multicolumn{1}{c}{$R$} & $\kappa_*$ & $\kappa$ &
 $\kappa_c$ & $\kappa_*/\kappa$ & $\tau$ \\
 & & \multicolumn{1}{c}{(kpc)} & & & & & \\
 & & & & & & & \\
\hline
 & & & & & & & \\
 HE1104-1805  & A &  7.88 & $0.050 \pm 0.019$ & 0.625 & 0.575 & 0.080 &
              $0.117 \pm 0.045$ \\
              & B & 15.29 & $0.014 \pm 0.005$ & 0.318 & 0.304 & 0.044 &
              $0.020 \pm 0.007$ \\
 HE0435-1223  & A &  7.52 & $0.037 \pm 0.014$ & 0.462 & 0.425 & 0.080 &
              $0.064 \pm 0.024$ \\
              & B &  6.77 & $0.043 \pm 0.016$ & 0.517 & 0.474 & 0.083 &
              $0.082 \pm 0.031$ \\
 RXJ1131-1231 & A &  9.37 & $0.017 \pm 0.006$ & 0.442 & 0.425 & 0.038 &
              $0.029 \pm 0.010$ \\
              & C &  9.52 & $0.016 \pm 0.006$ & 0.422 & 0.406 & 0.038 &
              $0.027 \pm 0.010$ \\
 WFI2033-4723 & B & 10.23 & $0.031 \pm 0.012$ & 0.380 & 0.349 & 0.082 &
              $0.047 \pm 0.019$ \\
              & C &  7.44 & $0.055 \pm 0.021$ & 0.610 & 0.555 & 0.090 &
              $0.123 \pm 0.060$ \\
 & & & & & & & \\
\hline	 
\end{tabular}
\end{table*}

To convert the surface brightness into surface mass density, $K$ and
evolutionary corrections from \cite{p97} were applied to the surface
brightness measures to provide a zero redshift galaxy profile in units of
stellar luminosity in the $H$-band.  To convert this into surface mass
density it is necessary to estimate the mass-to-light ratio of the stellar
population in the lensing galaxy.  Direct measurement of the stellar
mass-to-light ratio in galaxies is not easy due to the uncertain
contribution of dark matter, but for an early type galaxy typical of the
lensing galaxies in this sample, an appropriate mass-to-light ratio is
that measured for globular clusters with similar metallicity.
Observations of the mass-to-light ratio of globular clusters in the
Milky Way \citep{m05a} and M31 \citep{s11} are shown in Fig.~\ref{fig4},
plotted against metallicity.  The left hand panel is based on a catalogue
of dynamical masses for Milky Way globular clusters published by
\cite{m05a}, who use $V$-band photometry to derive values of $M/L_V$ for
the clusters in their sample.  Some 30 of these objects have been observed
in the near infrared bands by \cite{c07}, who also list values for [Fe/H].
We have used this photometry to convert the values of $M/L_V$ to $M/L_H$,
which are shown plotted against [Fe/H] in Fig.~\ref{fig4}.  The data show
no significant trend with metallicity, and the weighted mean of the
globular cluster mass-to-light ratios is $M/L_H = 0.56 \pm 0.21$, where we
take the dispersion as a measure of uncertainty.

The right hand panel of Fig.~\ref{fig4} shows measures of $M/L_K$ for M31
globular clusters from \cite{s11}.  The mass-to-light ratio is calculated
from direct measures of velocity dispersion and near infrared photometry,
and the data show some trend with metallicity as pointed out by
\cite{s11}.  This is especially true for low values of [Fe/H] where the
data are relatively sparse.  The weighted mean for the mass-to-light ratio
is $M/L_K = 0.53 \pm 0.24$, which after a small correction to convert to
the $H$-band using data from \cite{c07} gives $M/L_H = 0.57 \pm 0.26$ for
the M31 globular clusters.  In spite of the difference in average
metallicity and the trend in [Fe/H] versus $M/L_K$ for the M31 globular
clusters, the weighted means for the two samples in Fig.~\ref{fig4} are
very similar.  On this basis we shall adopt a value of
$M/L_H = 0.56 \pm 0.21$ which we apply to the surface brightness profile
to evaluate $\kappa_*$ as a function of galactocentric distance.  Error
bars are derived from errors in the measurement of surface brightness.
Values of $\kappa_*$ at the image positions of the quasars listed in
Table~\ref{tab1} together with errors resulting from uncertainty in the
value of $M/L_H$ are given in Table~\ref{tab2}.  Also shown is $R$, the
distance of the quasar image from the centre of the lensing galaxy.

Although it is not feasible to make direct measurements of the
stellar mass-to-light ratio of early-type galaxies due to the presence
of dark matter, stellar population synthesis (SPS) models provide an
alternative approach to estimating it.  \cite{c12} analyse the spectra
of 34 elliptical galaxies combined with a Kroupa IMF to model the stellar
mass-to-light ratio in the $K$-band.  They obtain an average value of
$M/L_K = 0.70 \pm 0.11$, which converts to $M/L_H = 0.75 \pm 0.12$.  This
is a little larger than, but consistent with, the values obtained from
direct measurements of $M/L_H$ for the globular cluster samples.

\subsection{Dark matter halo}

To obtain the effective optical depth to microlensing for the stars in
the lensing galaxy, we also need values for $\kappa_c$, the surface mass
density in smoothly distributed dark matter, at the positions of the
quasar images.  For this purpose we start with the values of total
convergence $\kappa$ at the quasar image positions from the mass models
of \cite{m09} and \cite{p12}, based on the software of \cite{k01}.  They
use a singular isothermal sphere plus external shear model
(SIS+$\gamma_e$) to estimate the total values of projected matter density
$\kappa$ and shear $\gamma$ at the location of the images.  In the case of
WFI2033-4723 a small improvement was obtained to the values of $\kappa$
and $\gamma$ by including a faint companion in the model \citep{b11}.  We
have verified these models using Keeton's software, and find that the
values of $\kappa$ and $\gamma$ are relatively insensitive to the exact
dark matter profile of the lensing galaxy as is well illustrated by the
work of \cite{w17} in their modelling of HE0435-1223.  They use 12
different models for the mass profile, and find that the resulting values
for $\kappa$ and $\gamma$ only vary by a few percent.  $\kappa_c$ is then
the difference between the values of $\kappa$ from the mass models and
$\kappa_*$ from the stellar surface brightness measuremants.

The optical depth to microlensing $\tau$ for stars as a function of
galactocentric distance can now be calculated from Eq.~\ref{eqn1} for the
four quasar systems, and are shown as filled circles in Fig.~\ref{fig5}.
Also shown as a solid line is the dark matter contribution $\kappa_c$ from
the SIS+$\gamma_e$ profile.  The positions of the quasar images are
indicated by arrows, and the corresponding values of $\kappa_c$ and $\tau$
are given in Table~\ref{tab2}.  Errors on $\tau$ based on those for
$\kappa_*$ are also shown.  The resulting values of $\tau$ put the stellar
populations of the lensing galaxies in the low optical depth regime
\citep{k97,l97}, and imply a low probability that stars in the lensing
galaxies can produce the strong microlensing features seen in
Fig.~\ref{fig2}.

\section{Microlensing simulations}
\label{sec4}

\begin{figure*}
\centering
\begin{picture} (0,540) (253,-260)
\includegraphics[width=0.49\textwidth]{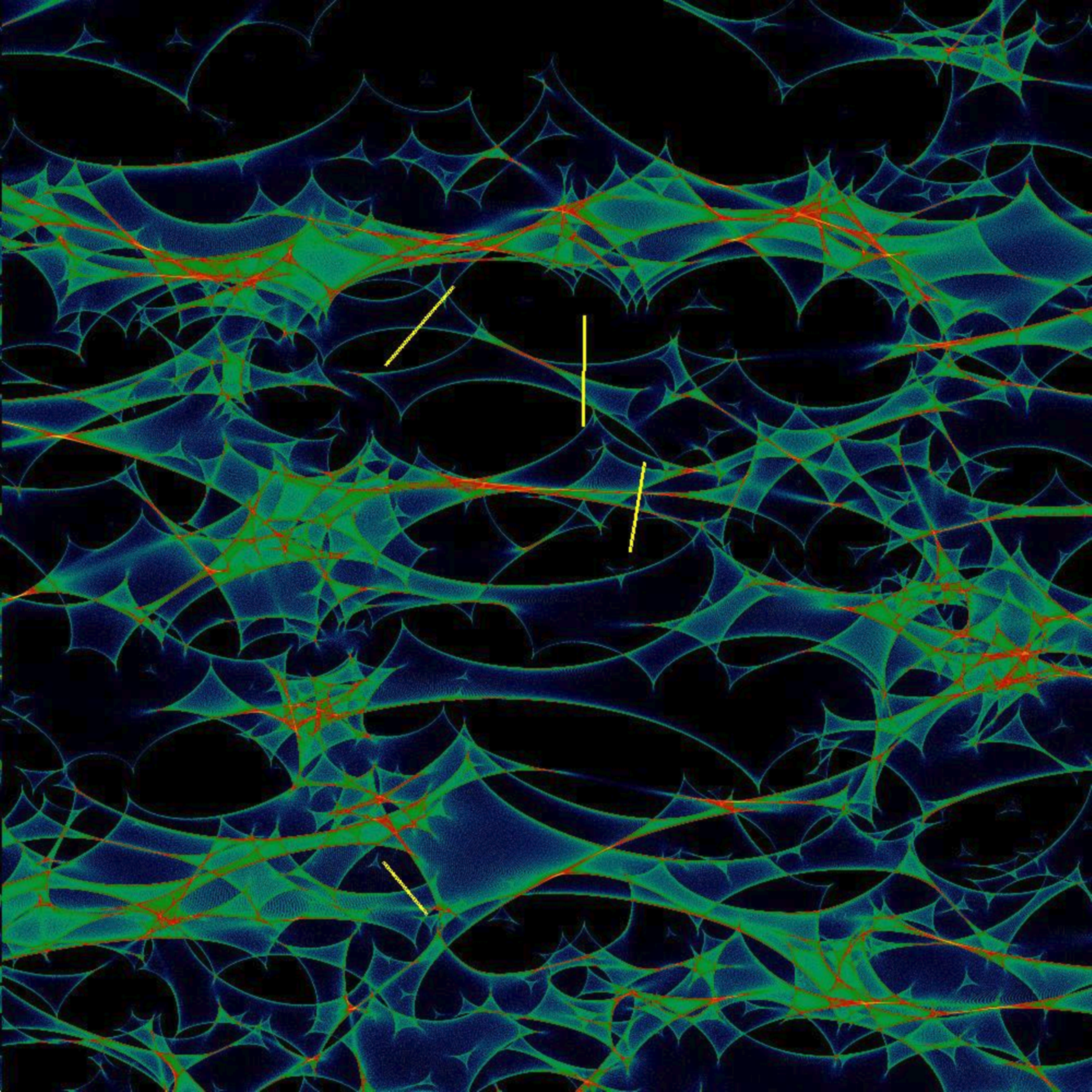}
\end{picture}
\begin{picture} (0,280) (-7,-260)
\includegraphics[width=0.49\textwidth]{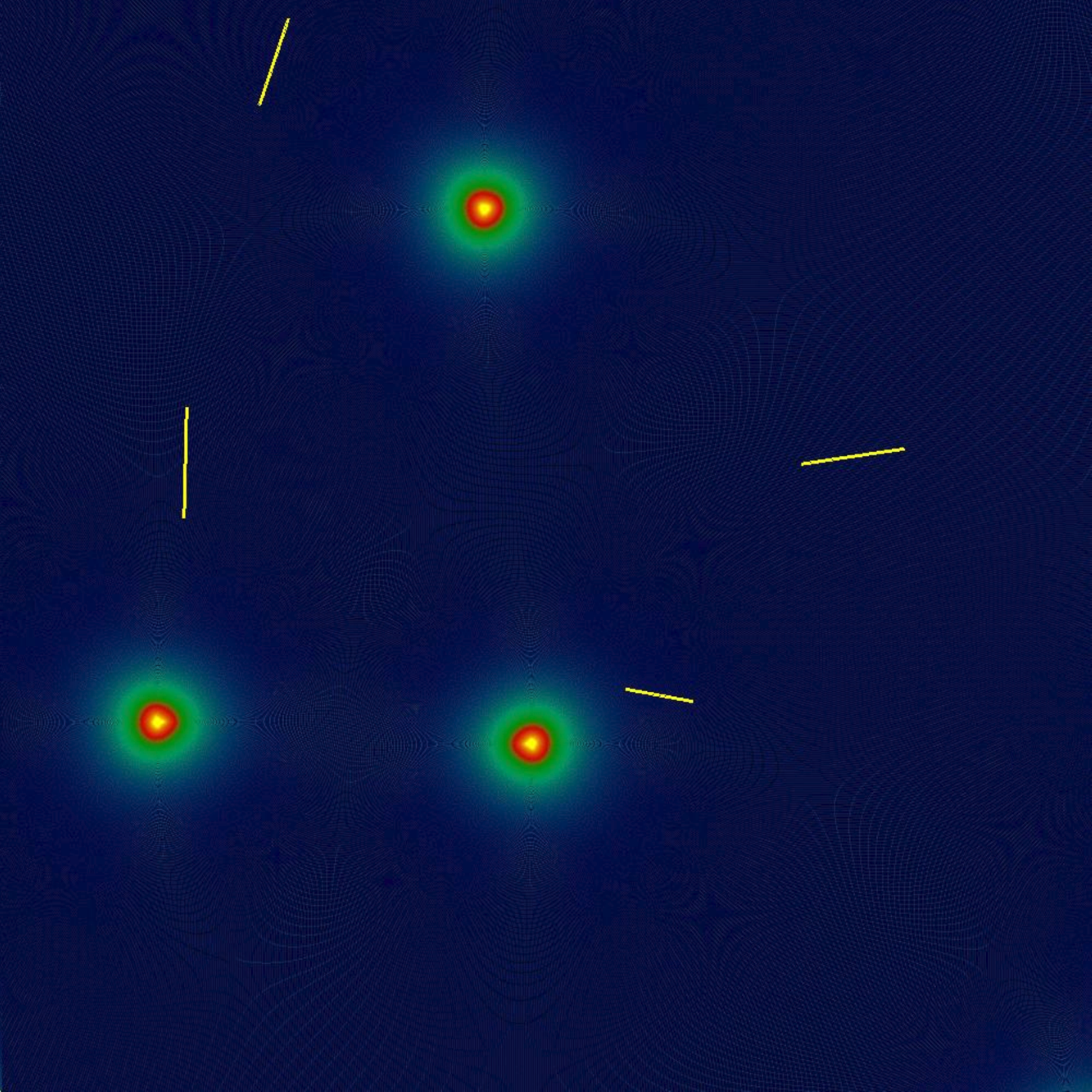}
\end{picture}
\begin{picture} (0,0) (260,3)
\includegraphics[width=0.49\textwidth]{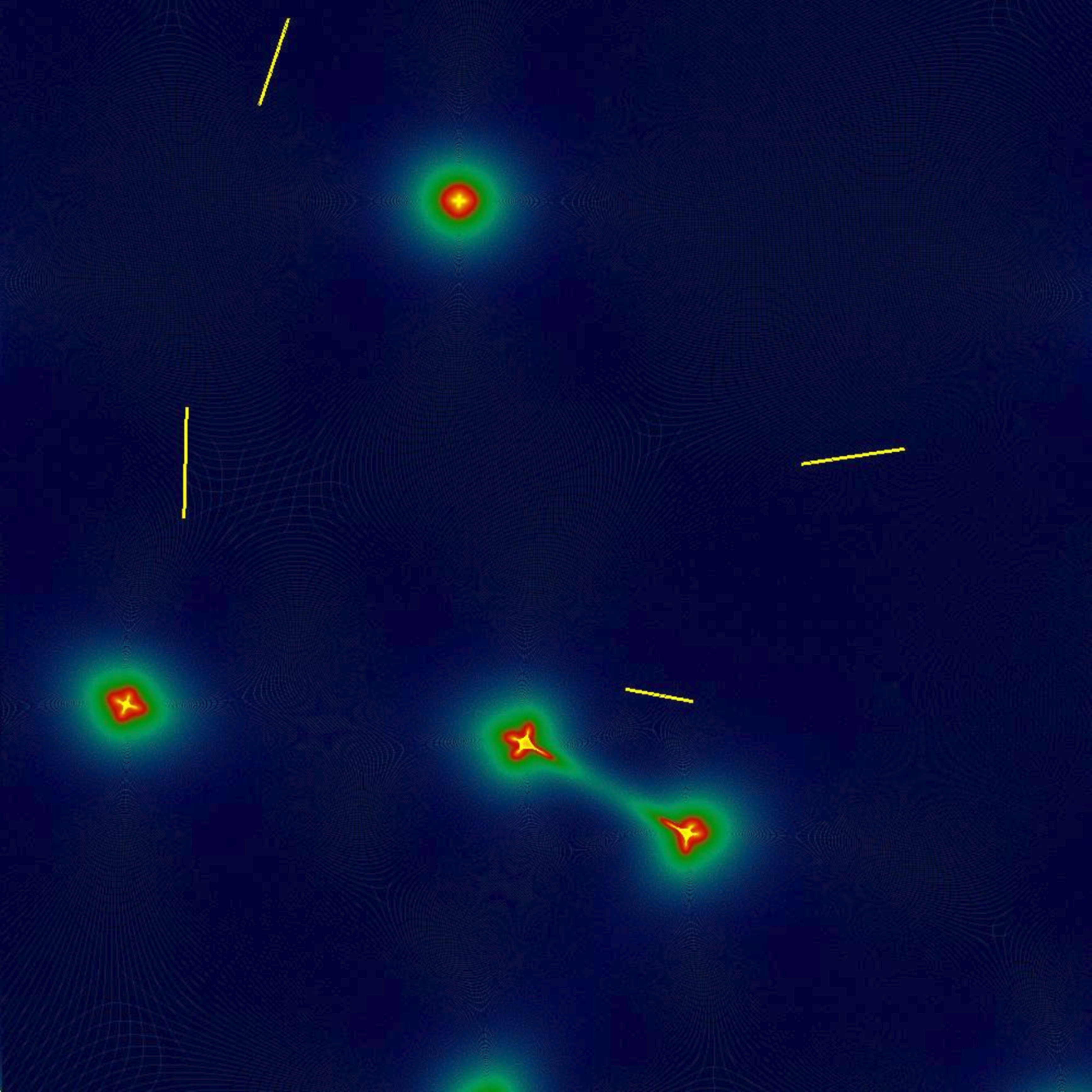}
\end{picture}
\begin{picture} (0,0) (0,3)
\includegraphics[width=0.49\textwidth]{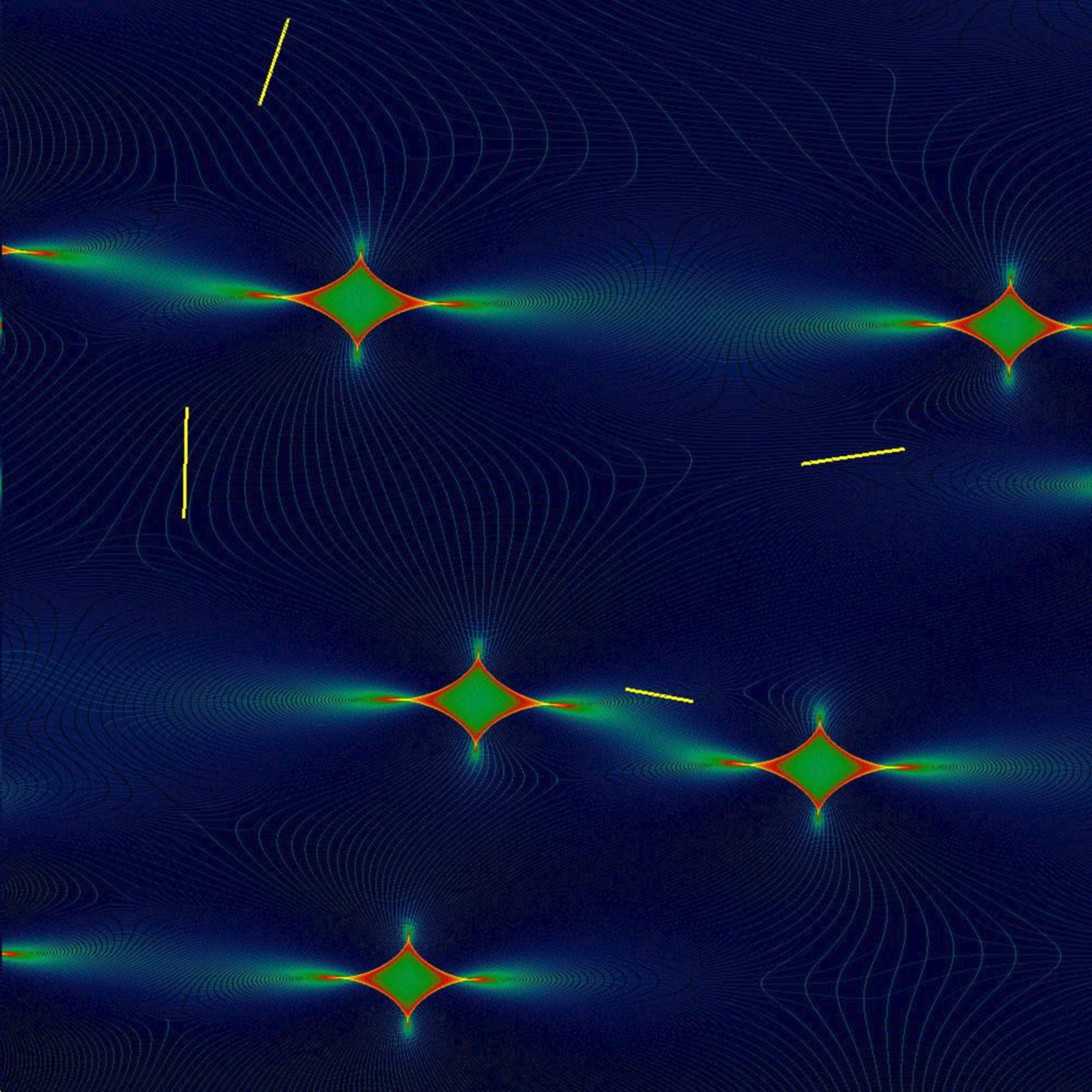}
\end{picture}
\caption{Microlensing magnification patterns for a population of 0.2
 M$_\odot$ bodies, assuming a point source.  The frames consist of
 $1000^2$ pixels and have a side length of 18 Einstein radii in the source
 plane.  The parameter values are $\kappa_* = 0.6$, $\kappa_c = 0$,
 $\gamma = 0.6$ (top left), $\kappa_* = 0.04$, $\kappa_c = 0$,
 $\gamma = 0$ (top right), $\kappa_* = 0.04$, $\kappa_c = 0.4$,
 $\gamma = 0$ (bottom left) and $\kappa_* = 0.04$, $\kappa_c = 0.4$,
 $\gamma = 0.4$ (bottom right).  Yellow lines indicate  tracks across the
 amplification pattern for the length of the four light curves in
 Fig.~\ref{fig2}, and a net transverse velocity of 600 km sec$^{-1}$ is
 assumed.  The position and orientation of the tracks is random.}
\label{fig6}
\end{figure*}

\begin{figure}
\centering
\begin{picture} (0,200) (130,0)
\includegraphics[width=0.5\textwidth]{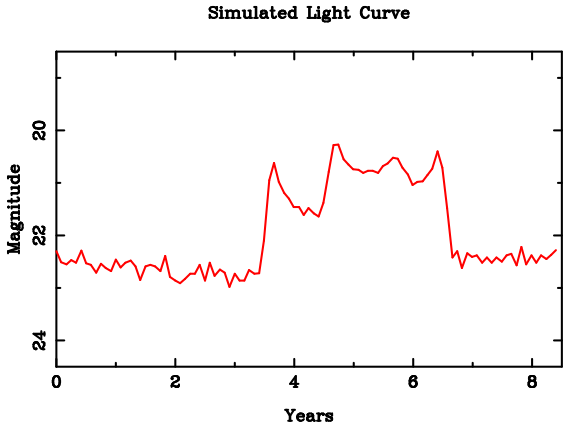}
\end{picture}
\caption{Light curve corresponding to the longest track in the top left
 panel of Fig.~ \ref{fig6}.}
\label{fig7}
\end{figure}

In this section we make use of microlensing simulations to
illustrate the optical depth regimes relevant to this paper, and then
estimate the probability that the amplitudes of the observed light curves
can be achieved by microlensing from simulated star fields appropriate to
the line of sight to the quasar images.  The first studies of microlensing
in gravitationally lensed quasar systems were based on the quadruple
Q2237+0305 \citep{i89,w90}, also known as the Einstein Cross.  Due to the
overall configuration of this system, and in particular the very low
redshift of the lensing galaxy ($z = 0.0394$) the four quasar images lie
within a kiloparsec of the galaxy centre.  This results in a high optical
depth to microlensing for the stellar population ($\tau \gtrsim 0.6$),
which proved very useful for microlensing studies \citep{w00,a08}.  These
values of $\tau$ are in the high optical depth regime defined by
\cite{k97}, where strong caustic-induced features dominate the
magnification pattern.  The top left panel of Fig.~\ref{fig6} shows a
simulated magnification pattern for a point source with stellar surface
density $\kappa_* = 0.6$ and $\gamma = 0.6$, similar to that found for
Q2237+0305 \citep{w00}.  The simulation was carried out using the code of
\cite{w99}, freely available on the internet.  The yellow lines show
tracks in the source plane of the same length as the light curves in
Fig.~\ref{fig2}, using redshifts for the systems from Table~\ref{tab1},
and assuming a transverse velocity of 600 km sec$^{-1}$.  The tracks have
random position and orientation, and illustrate the high probability that
quasar images are microlensed by such a large optical depth of lenses.
The red line in Fig.~\ref{fig7} shows the light curve corresponding to the
longest track in the top left panel of Fig.~\ref{fig6}, and it is
seen that the image fluctuates in brightness by more than two magnitudes.

\begin{figure*}
\centering
\begin{picture} (0,280) (255,0)
\includegraphics[width=0.49\textwidth]{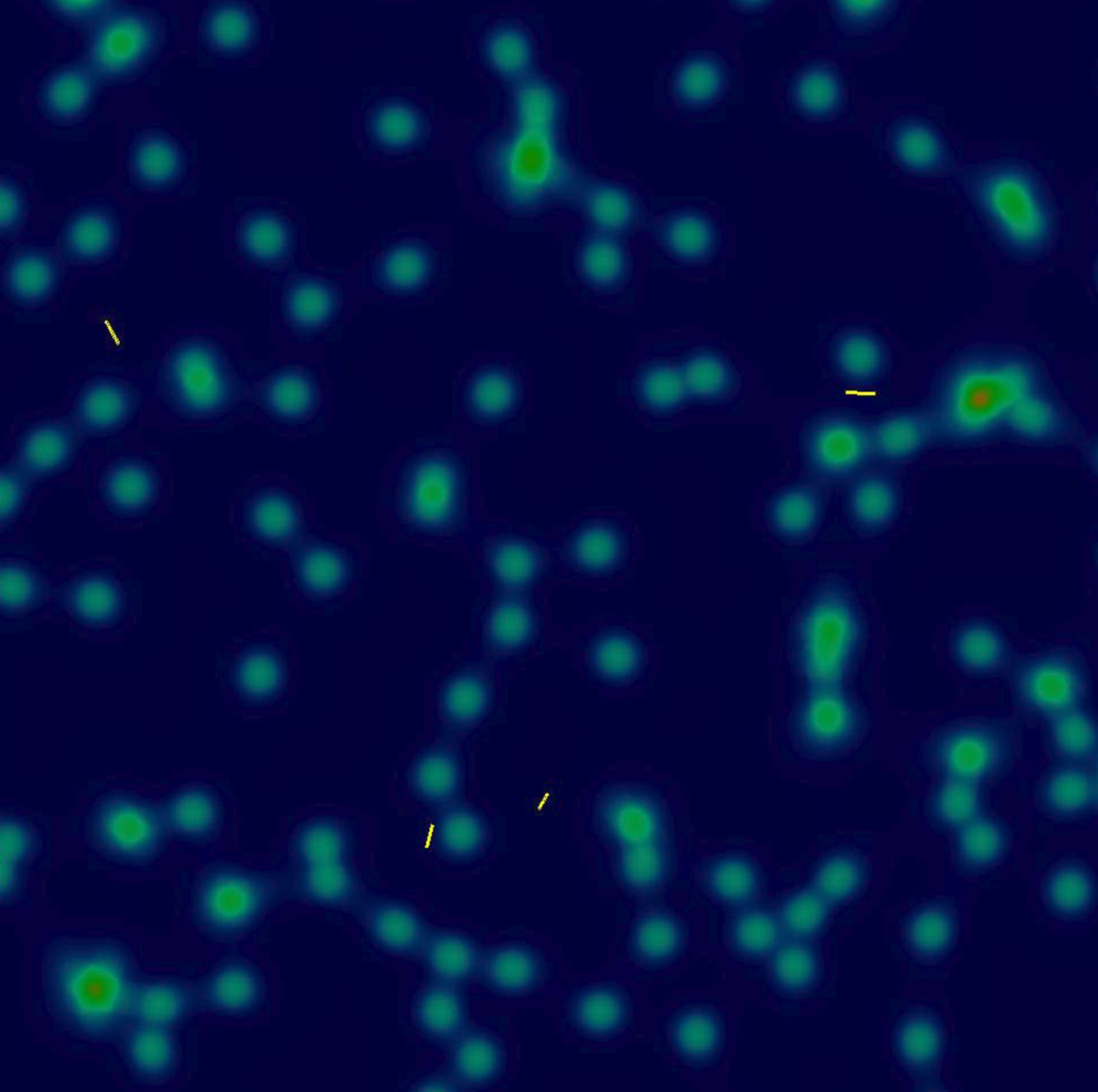}
\end{picture}
\begin{picture} (0,280) (-5,0)
\includegraphics[width=0.49\textwidth]{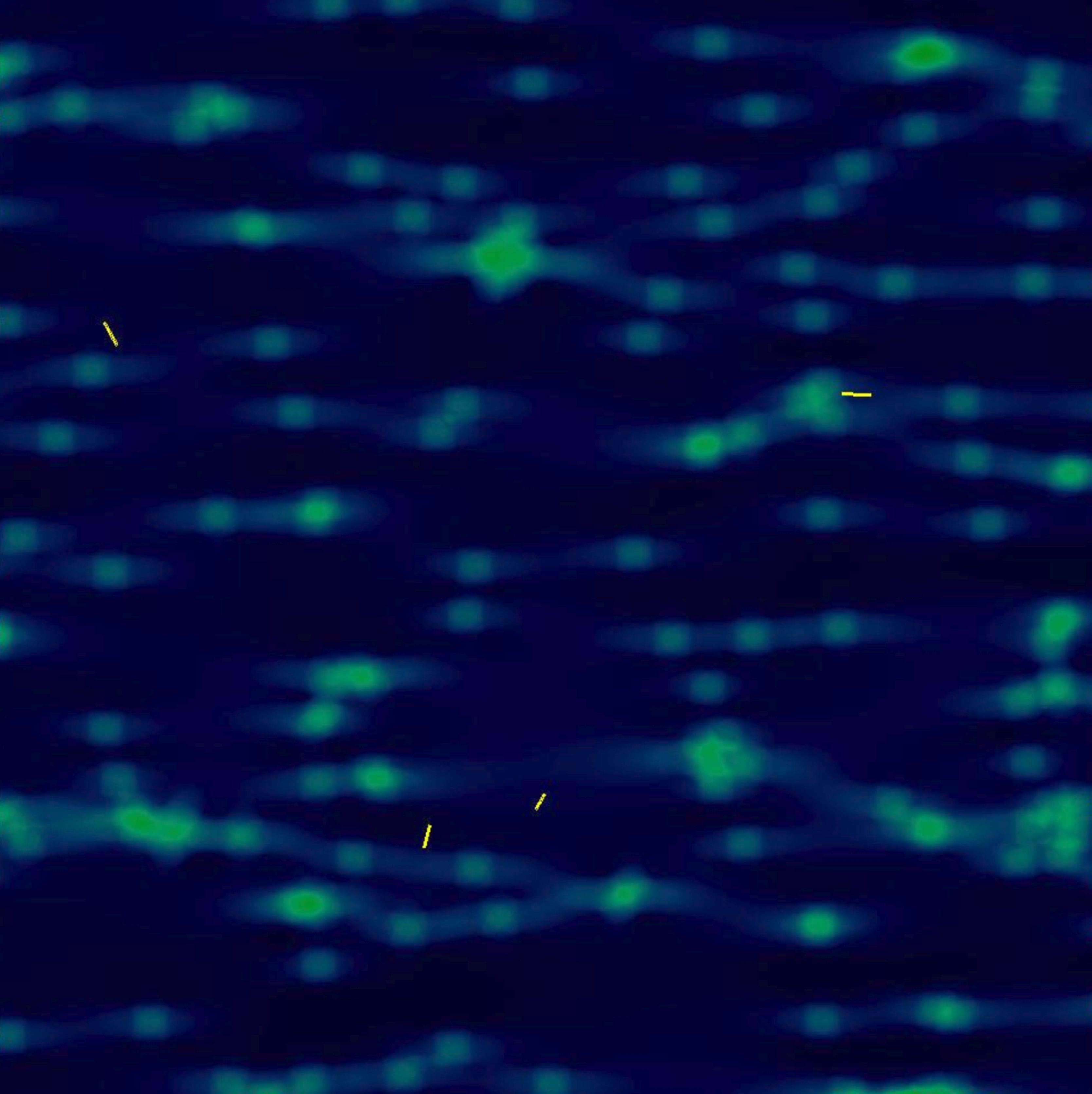}
\end{picture}
\caption{Microlensing magnification patterns for a population of 0.2
 M$_\odot$ bodies with parameter values $\kappa_* = 0.04$,
 $\kappa_c = 0.4$, $\gamma = 0$ (left hand panel) and $\kappa_* = 0.04,
 \kappa_c = 0.4$, $\gamma = 0.4$ (right hand panel).  Each frame consists
 of $1000^2$ pixels and has a side length of 72 Einstein radii in the lens
 plane.  The tracks are as for Fig.~\ref{fig6}.  The plots have been
 convolved with a source size of $0.5 R_E$.}
\label{fig8}
\end{figure*}

The top right hand panel of Fig.~\ref{fig6} shows a simulated
magnification pattern for a population of 0.2 M$_\odot$ compact bodies
with a surface density of $\kappa_* = 0.04$, typical of the values for the
quasar sample in Table~\ref{tab2}.  These parameters put the value of
$\tau$ in the low optical depth regime \citep{k97}, where strong
caustic-induced features do not form.  The simulation assumes a point
source to highlight the fine structure of the magnification pattern, and
shows well-separated lenses with clearly defined impact parameters and
little sign of distortion.  The bottom left hand panel shows the effect of
the presence of smoothly distributed matter with $\kappa_c = 0.4$,
similar to the values for the dark matter halos in Table~\ref{tab2}.  The
addition of dark matter to the simulations results in the appearance of
new images, in line with the value of $\tau$ from Eq.~\ref{eqn1}, but
there is still little sign of caustic features.  The bottom right hand
panel shows the effect of including a shear term in the simulation, equal
in size to $\kappa_c$.  It is seen that the magnification
pattern is smeared out in the direction of the shear, reducing the maximum
magnification somewhat.  Similar tracks to those in the top left hand
panel are superimposed on the amplification patterns to provide a visual
illustration of the implications of the low values of $\tau$ from
Table~\ref{tab2}, and the likelihood of a microlensing event in the
observed light curves.  All the tracks lie well clear of any lensing
feature, and the fluctuations in their light curves are negligible.

The probability that the quasars in the sample are microlensed with the
observed amplitude difference by stars in the lensing galaxies can be
estimated by enlarging the simulated area to contain more lenses, and
using the parameters for $\kappa_*$ and $\kappa_c$ from Table~\ref{tab2}
to provide a simulation for each of the eight quasar images.  As an
illustrative example, the left hand panel of Fig.~\ref{fig8} shows a
simulated magnification pattern for an area 16 times larger in the source
plane than for Fig.~\ref{fig6}, using typical values for $\kappa_*$ and
$\kappa_c$ from Table~\ref{tab2}.  To enable a realistic estimate of the
probability of microlensing we must take into account the finite size of
the continuum disc of the quasar.  \cite{j15a} use a microlensing study of
a large sample of lensed quasars to obtain a value of 7.9 light days for
the half-light radius of the accretion disc at 1736\AA\ for an average
quasar.  This figure is a factor of two larger than their earlier result
\citep{j12} and consistent with the compilation by \cite{m13}.  It also
appears to be consistent with the reverberation mapping results of
\cite{m18} for quasars.  Given the sensitivity of the microlensing
amplitude to the source size we shall conservatively adopt the $2\sigma$
lower bound of 4 lt-day or $10^{16}$ cm from \cite{j15a}.  Combining this
with values for the Einstein radius $R_E$ from Table~\ref{tab1} gives a
disc radius of around $0.5 R_E$ which we use in the simulation in
Fig.~\ref{fig8}.  The sensitivity of our results to source size is
discussed further in Section~\ref{sec8}.  Yellow tracks corresponding to
the length of the light curves in Fig.~\ref{fig2}, but in random position
and orientation, are also plotted as for Fig.~\ref{fig6}.  The simulation
in the right hand panel of Fig.~\ref{fig8} is as for the left hand panel,
but with the addition of a shear term equal to $\kappa_c$.  The effect of
shear is basically to smear the magnification pattern along a preferred
axis in such a way that small magnifications become slightly more
frequent, and high magnifications are less likely to be achieved.  As
\cite{s92} point out, in many applications the effect of the shear is
neglected for simplicity, but we include the right hand panel of
Fig.~\ref{fig8} to clarify the point.

\begin{table*}
\caption{Microlensing probabilities}
\label{tab3}
\begin{center}
\vspace{5mm}
\begin{tabular}{lcccccccrc}
\hline\hline
 & & & & & & & & & \\
 Quasar System & \multicolumn{3} {c} {Track 1} &
 \multicolumn{3} {c} {Track 2} & Trials &
 $N(\Delta\!>\!\Delta m)$ & $P(\Delta m)$ \\
 & $\kappa_*$ & $\kappa_c$ & $\gamma$ & $\kappa_*$ &
 $\kappa_c$ & $\gamma$ & & & \\
 & & & & & & & & & \\
\hline
 & & & & & & & & & \\
 HE1104-1805  & 0.050 & 0.575 & 0.000 & 0.014 & 0.304 & 0.000 &
 $10^6$ &  41013 & 0.041 \\
              & 0.050 & 0.575 & 0.493 & 0.014 & 0.304 & 0.200 &
 $10^6$ &  46445 & 0.046 \\
 & & & & & & & & & \\
 HE0435-1223  & 0.037 & 0.425 & 0.000 & 0.043 & 0.474 & 0.000 &
 $10^6$ &  40003 & 0.040 \\
              & 0.037 & 0.425 & 0.393 & 0.043 & 0.474 & 0.593 &
 $10^6$ &  49920 & 0.050 \\
 & & & & & & & & & \\
 RXJ1131-1231 & 0.017 & 0.425 & 0.000 & 0.016 & 0.406 & 0.000 &
 $10^6$ &     98 & 0.000 \\
              & 0.017 & 0.425 & 0.597 & 0.016 & 0.406 & 0.504 &
 $10^6$ &      6 & 0.000 \\
 & & & & & & & & & \\
 WFI2033-4723 & 0.031 & 0.349 & 0.000 & 0.055 & 0.555 & 0.000 &
 $10^6$ &  63074 & 0.063 \\
              & 0.031 & 0.349 & 0.250 & 0.055 & 0.555 & 0.730 &
 $10^6$ & 131042 & 0.131 \\
 & & & & & & & & & \\
\hline	 
\end{tabular}
\end{center}
\end{table*}

For each of the quasars in Table~\ref{tab1} two simulations were run, one
for each image, with appropriate values for $\kappa_*$ and $\kappa_c$.
In the first instance $\kappa_*$ and $\kappa_c$ were set equal to
the values given for the two corresponding images in Table~\ref{tab2},
together with a source of half-light radius 4 lt-day.  Tracks
corresponding to the length of the light curve for that quasar were
placed on the two simulations in random positions and orientations, and
the magnitudes from the amplification patterns subtracted to give the
difference light curve.  The amplitude $\Delta$ of this difference light
curve was then compared with the observed amplitude $\Delta m$ from
Table~\ref{tab1}.  By repeating this procedure a number of times, the
frequency with which $\Delta$ exceeded $\Delta m$ was counted, and hence
the probability that the observed value of $\Delta m$ was the result of
microlensing by the simulated star field corresponding to the lensing
galaxy was estimated.  The results of this process are given in the first
line of data for each quasar in Table~\ref{tab3}.  The first seven columns
identify the quasar system and give the parameters used in the simulation
for each of the two associated images.  Subsequent columns give the number
of runs with random placing of the light curves, the number of times
$N(\Delta>\Delta m)$ that the amplitude of the simulated difference light
curve exceeded the observed amplitude, and the consequent probability
$P(\Delta m)$ that it is the result of microlensing.  Due to the large
number of trials, the Poisson error on this probability is quite small,
less than 0.0003 in all cases.  The second lines of data in
Table~\ref{tab3} show the results for simulations with the same
parameters as for the first line, but with the addition of a shear term
from the mass models used to estimate the convergence at the image
positions.

In order to estimate the probability that all four quasar systems would be
found to be microlensed as observed, we combined the four sets of
simulations in Table~\ref{tab3} and counted the number of times that all
four quasars simultaneously exceeded the amplitude of their light curves
in Fig.~\ref{fig2}.  This figure was found to be zero, with or without a
shear contribution.  Given that RXJ1131-1231 was observed to vary with a
very large amplitude which may be a statistical fluke, we also counted the
number of times that all but one of the quasars exceeded their observed
amplitude.  In this case for the runs with shear set to zero, and for the
non-zero values of $\gamma$ from Table~\ref{tab3}, the numbers were 111
and 318 respectively.  We discuss the significance of these results in
Section~\ref{sec5}.

\section{Results}
\label{sec5}

The results of this paper are basically summarised in Tables~\ref{tab2}
and \ref{tab3}, and the data in each table may be used to estimate the
probability that the microlensing observed in the four quasar systems
in Table~\ref{tab1} can be attributed to stars in the lensing galaxies.
To put it another way, we take as a statistical hypothesis that the
observed microlensing in at least one quasar system is not caused by stars
in the lensing galaxy.  This is the question of interest, as it measures
the probability that there is some other population of compact bodies
along the line of sight.  In Table~\ref{tab2} the values for $\tau$ are in
the low optical depth regime and effectively measure the probability of
microlensing by a star in the lensing galaxy.  Given that the light path
to each quasar image is different, and hence the values of $\tau$ are
independent, they can be combined in a straightforward way to give the
probability that all four systems are being simultaneously microlensed by
stars in the lensing galaxies, or equivalently that at least one system is
not.  Using the data for the images which become brighter gives a
probability of $1 \times 10^{-5}$ that all the quasars are microlensed by
stars.  As we discuss below, there is a case to be made that RXJ1131-1231
is in some way untypical of lensing systems due to its very low value for
$\tau$.  If we allow RXJ1131-1231 to be excluded by requiring only 3
systems to be microlensed, this increases the probability to
$3 \times 10^{-4}$.

The results from the microlensing simulations in Table~\ref{tab3} provide
an alternative way of estimating the probability that a stellar
distribution similar to the one seen in the lensing galaxies can produce
the observed microlensing.  For each quasar system the first line in
Table~\ref{tab3} shows microlensing simulations using values for
convergence from lenses $\kappa_*$ and smoothly distributed matter
$\kappa_c$ equal to the values in Table~\ref{tab2} for the lensing
galaxies at the positions of the quasar images.  The source size was set
to 4 lt-day, as discussed in Section~\ref{sec4}.  As might be expected,
light curves with larger amplitudes are less likely to be produced in the
simulations as can be seen in Table~\ref{tab3}, especially for 
RXJ1131-1231.  If RXJ1131-1231 is included in a probability calculation as
for Table~\ref{tab2}, the probabilty of all systems being microlensed by
stars is neglibly small ($< 10^{-6}$).  However, there is some reason to
believe that RXJ1131-1231 may have an unusually large amplitude
\citep{p12}, and so to make a more conservative estimate of the
probability that the stars are responsible for the microlensing we have
repeated the calculation excluding RXJ1131-1231.  In this case the
probability that all three of the remaining systems are microlensed by
the stars is $1 \times 10^{-4}$, of the same order as the result for the
data in Table~\ref{tab2}.  For the simulations including shear terms, the
probability increases to $3 \times 10^{-4}$.  This is close to the result
of the simulation counting the number of times at least 3 systems were
simultaneously microlensed, where the 318 coincidences in $10^6$ trials
give a probability of $3 \times 10^{-4}$.

RXJ1131-1231 has stood out in our analysis as anomolous, with an
exceptionally small value of $\tau$ and a large change in brightness,
resulting in a very low probability for microlensing from the observed
stellar population.  This system has already been identified by \cite{p12}
as an outlier in their analysis, and we feel it safest to exclude it from
our main result, as its inclusion heavily weights our conclusion against
the possibility that the stellar population can cause the observed
microlensing.  The most obvious explanation is that the lensing galaxy
contains a large component of dark microlensing mass, but to follow this
up would be beyond the scope of the present paper.

\section{Comparison with other work}
\label{sec6}

An alternative approach to estimating the likelihood that stars in the
lensing galaxies are responsible for the observed microlensing comes from
simulating microlensing patterns for a range of values for the fraction
of mass in compact objects.  A maximum likelihood or Bayesian analysis
is then used to determine the most likely mass fraction to be responsible
for the observed microlensing amplitude \citep{m09,p12}.  These two
studies find a most likely overall mass fraction in compact objects
to replicate the observed microlensing of less than 10\%, which in
general would be compatible with the size of the stellar population of the
lensing galaxies.  For comparison with the analysis in this paper we have
listed our values for the mass fraction in stars at the quasar image
locations in Table~\ref{tab2}.  They are quite similar to the values from
\cite{m09} and \cite{p12}, and in fact if they are plotted against the
the data in Fig. 6 of \cite{j15b} they follow the line of their composite
model with a small negative offset of about 20\%.  From this we deduce
that any difference in our conclusions is not due to differing values for
the mass fraction, but depends on whether the resulting surface density of
stars can actually produce the observed microlensing.  As has been
discussed in Section~\ref{sec4}, microlensing amplitudes are strongly
dependent on the size of the quasar source.  \cite{m09} assume a source
size of ~1 lt-day, which is much less than the value we have used and
could well account, at least in part, for the apparent difference in our
results.  \cite{p12} base their analysis on X-ray microlensing, and feel
justified in assuming a point source.  In this case it is not so easy to
make a direct comparison with our results as there is considerable
uncertainty as to the actual size of the X-ray emitting region.

It is important to note that the samples of quasar systems used in the
two approaches are very different.  The samples chosen by \cite{m09} and
\cite{p12} are much larger than the one in this paper, and mostly composed
of small separation systems where the quasar images are deeply embedded in
the lensing galaxies, and can plausibly be microlensed by stars.  By
contrast, the sample we use here has been specifically chosen so that
the quasar images lie clear of the bulk of the stellar distribution.  The
significance of this is made clear in Fig. 12 of \cite{p12} where wide
separation lenses (including most of our sample) have a very different
probability distribution to their sample as a whole, with a much higher
likelihood of a dark matter component consisting of compact objects.

Looking at the probability distributions in more detail raises some
interesting questions.  In Fig. 7 of \cite{p12} the histograms for most
objects are quite flat, with probabilities in each bin ranging from about
5\% to 10\%.  An interesting exception is RXJ1131-1231 which shows a
steady increase in probability to high values of stellar mass fraction.
The histogram for HE0435-1223 is not easy to interpret, as it shows two
almost equal maxima at 2\% and 100\% stellar mass fraction.

It is worth emphasising that the Bayesian analysis and the procedure in
this paper are essentially measuring different things.  The question we
have addressed is the likelihood that microlensing by the observed
stellar population can always reproduce features of the observed
difference light curves, and this means that the stellar component is not
treated as a free parameter.  By contrast, the Bayesian analysis is
measuring the most likely mass fraction in compact bodies.  Given that the
two samples of lensed systems are quite different, an extensive catalogue
of mostly compact systems compared with the small number of known wide
separation systems, it is perhaps to be expected that the results are
different, but not necessarily contradictory.

\section{The identity of the lenses}
\label{sec7}

The idea that dark matter is in the form of compact bodies has a long
history.  In an early review of dark matter, \cite{t87} considered such
bodies as plausible candidates alongside various elementary particles.
\cite{h93} proposed that there was observational evidence for a
cosmological distribution of compact bodies which betrayed their
presence by microlensing quasars, and the evidence for this has recently
been reviewed and consolidated \citep{h11}.  The case for compact bodies
as dark matter received a severe setback with the publication of results
from microlensing surveys in the Galactic halo \citep{a00,t07,w11}, which
put limits on any population of solar mass compact bodies.  However, more
recent measures of the structure of the Galactic halo, and reassessment of
other aspects of the surveys, have now thrown doubt on the reliability of
the limits on dark matter in the form of compact bodies set by these
groups \citep{h15,g17,c18}.

If dark matter is indeed made up of of compact bodies, then there are
several strong constraints on the form it can take.  Combining
measurements of light element abundances with the photon density from the
CMB gives a fairly robust limit of $\Omega_b < 0.03$ \citep{s93} which
effectively rules out baryonic objects as dark matter candidates.  This of
course includes stars and planets as well as later products of stellar
evolution such as white dwarfs, neutron stars and stellar black holes.
A second constraint comes from the timescale of microlensing events, which
constitute much of the evidence for compact bodies as dark matter.  Both
the MACHO events in the galactic halo and the microlensing events observed
in multiple quasar systems of the type discussed here point to a lens mass
of around a solar mass, with an uncertainty of at least a factor of 10
either way.  This is nonetheless sufficient to rule out a wide range of
candidates, which we discuss below.  A final constraint is compactness.
Again, since much of the evidence for a cosmological distribution of
compact bodies comes from microlensing, dark matter bodies must be compact
enough to act as lenses in the configuration in which they are observed.

In her review of the nature of dark matter, \cite{t87} lists quark
nuggets, cosmic strings and primordial black holes as non-baryonic
non-particle candidates.  Quark nuggets have some virtue as dark matter
candidates \citep{a99}, but as their mass is limited to around
$10^{-8}$ M$_\odot$ they cannot be the objects detected in the microlensing
experiments.  The possibility that cosmic strings might betray their
presence by the microlensing of distant quasars has been investigated in
some detail by \cite{k08}.  Their motivation is to establish a method for
detecting the presence of cosmic strings, and they estimate that a typical
microlensing event would last about 20 years.  This fits in well with
other microlensing signatures, but the problem they face is that such
events would be extremely rare.  The optical depth to microlensing for
cosmic strings derived by \cite{k08} is of the order $\tau \sim 10^{-8}$
which completely rules them out as the source of the microlensing seen in
multiple quasar systems.

An interesting idea proposed by \cite{w98} suggesting that extreme radio
scattering events might be caused by cold self-gravitating gas clouds has
been investigated by \cite{r01} with a view to establishing whether such
clouds could be detected by microlensing searches.  The clouds are
predicted to have masses of around $10^{-3}$ M$_\odot$ and radii of
$\sim10$ AU, and would make up a significant fraction of the mass in
galaxy halos.  \cite{r01} concluded that although strongly constrained by
MACHO microlensing searches, such clouds could not be definitively ruled
out as the dark matter component of the Galaxy halo.  However, as an
explanation for the microlensing seen in multiple quasar systems they
can be ruled out on the basis of their low proposed mass, and also the
fact that they are essentially baryonic and so cannot make up the dark
matter on cosmological scales.

The steady improvement in numerical simulations of the formation of
large-scale structure from primordial perturbations culminating in the
discovery of a universal profile for dark matter halos \citep{n96} has
raised the question of the nature of their internal strucure on small
scales.  An ambitious programme to solve this problem was undertaken
by the Aquarius Project \citep{s08} which focussed on simulating the
evolution of individual galaxy halos with unprecedented resolution.
The two main results of this and other similar simulations were that
galaxy halos have cuspy dark matter profiles, and that there should be a
very large population of dark matter subhalos.  At first sight these
subhalos might seem attractive candidates for microlensing quasars, but
their structure as described by \cite{s08} imply that they would not be
compact enough to act as microlenses for the accretion discs of distant
quasars, even if subhalos as small as a solar mass can form.

The objects which have received the most attention as candidates for dark
matter in the form of compact bodies are primordial black holes.  They
satisfy the basic constraints mentioned above in that they are
non-baryonic, very compact and can in principle form over a wide range of
masses, including those relevant to quasar microlensing.  The recent
detection of gravitational waves from a black hole merger \citep{a16}
also supports the idea that dark matter may be in the form of primordial
black holes, although there is still debate as to whether the
gravitational wave detections should be attributed to black holes of
stellar origin.  The question of whether primordial black holes in the
right mass range can actually be created is the subject of ongoing debate
(see for example \cite{c15} and references therein), but as things stand
they appear to be the only plausible candidates for dark matter in the
form of compact bodies.  Additional support for this conclusion comes from
recent work \citep{b18} using new lattice computations to give an accurate
QCD equation of state.  The softening of the equation of state during the
QCD phase transition implies an enhancement in the production of
primordial black holes peaking at around $0.7 M_\odot$.

\section{Discussion}
\label{sec8}

This paper sets out to examine the assumption that the microlensing
observed in multiple quasar systems can always be accounted for by the
stellar population of the lensing galaxy.  The investigation was prompted
by the observation that in some such systems the quasar images appear to
lie well clear of the lensing galaxy, raising the question of the
probability of them being microlensed by the lensing galaxy stars.  To
assess this probability, the four systems analysed here were chosen
specifically because the quasar images were distant from the lensing
galaxy.  As for nearly all such systems which have been adequately
monitored, the individual quasar images were strongly microlensed.  The
calculation of the optical depth to microlensing was not modelled, but
based on direct measurement of the starlight, and hence of the surface
density of stars along the line of sight, together with values for total
convergence from lens models.  The surface brightness measures were made
in the near infrared $H$-band to optimise the contrast between galaxy and
quasar light.  Measurement of the surface brightness profiles of the
lensing galaxies was made more complicated by the distorting effects of
the superimposed quasar images, but in practice it turned out to be quite
easy to remove the contribution of the star-like quasar images to obtain a
good estimate of the underlying galaxy surface brightness.  After careful
normalisation, clean featureless profiles were obtained which in all cases
closely followed the de Vaucouleurs $r^{1/4}$ law. The good agreement with
the work of \cite{b97} illustrated in Fig.~\ref{fig3} gives further
confidence that the surface brightness profiles are fair measures of the
starlight in the lensing galaxies.

One of the advantages of working in the near infrared $H$-band is the
small dispersion in the stellar mass-to-light ratio ($M_*/L_H$) for
different stellar populations. The pervasive presence of dark matter
severely limits the systems which are suitable for direct measurements
of $M_*/L_H$, and effectively confines the candidates to globular
clusters.  We have used samples of globular clusters from the
Milky Way and M31 to estimate the stellar mass-to-light ratio of our
sample of lensing galaxies.  Despite significant differences in the
relationship between [Fe/H] and $M/L_H$ for the two samples, the mean
and dispersion of $M/L_H$ were very similar.  The reliability with which
these values of mass-to-light ratio can be used to estimate $M/L_H$ for
the lensing galaxies is supported by the evidence for a universal IMF.
In a comprehensive review covering a wide range of stellar environments
from local stellar populations to distant galaxies \cite{b10} conclude
that `there is no clear evidence that the IMF varies strongly and
systematically'.  There have nonetheless been claims of evidence for a
`bottom heavy' IMF in massive early-type galaxies \citep{v10} where the
strength of spectral lines associated with low mass stars in a small
sample of elliptical galaxies were argued to be inconsistent with a
universal IMF.  This conclusion is tentatively supported by \cite{t10}
from a study of 56 early-type galaxies using a variety of techniques to
obtain the absolute normalisation of the IMF.   However, \cite{s15a}
showed that not all such galaxies have IMFs differing significantly from
that of the Milky Way.

These inconsistencies seem to have been largely removed with the
availabilty of high spatial resolution spectra, which appear to show a
deficit of low mass stars in the central cores of some giant elliptical
galaxies \citep{m15}.  This results in an atypical IMF in the galactic
centre, for reasons which are still not properly understood, but
throughout the rest of the galaxy the IMF is consistent with that of the
Milky Way.  This then supports the use of a universal IMF in the
calculation of the stellar surface density at the positions of the lensed
quasar images which are invariably well clear of the core regions of the
lensing galaxies.  In addition, recent work by \cite{s15b} has provided an
explanation for the small values of $M/L_K$ at high metallicities, and
confirms the value we have adopted from Fig.~\ref{fig4}.  Further support
for a low mass-to-light ratio in the near infrared is presented by
\cite{i13}, who show that taking into account the role of the asymptotic
giant branch results in lower estimates of stellar mass in galaxies.

An alternative approach to estimating the mass-to-light ratios of galaxies
comes from SPS modelling of stellar populations.  The values of $M/L_H$
derived from the models of \cite{c12} applied to their observations of 34
late-type galaxies are consistent with the dynamically based values from
globular clusters in the Galaxy and M31.  The small discrepancies between
these two independent methods of estimating the mass-to-light ratio are
not large enough to significantly affect the conclusions of this paper.

It is seen from Table~\ref{tab2} that the probability for any
individual quasar image in our sample to be microlensed by a star is in
the range 2\%-12\%.  The uncertainties in these figures come from three
different sources.  The first of these is the assumed value for the
$H$-band mass-to-light ratio.  The errors given in Table~\ref{tab2} are
based on dispersion in large samples of Galactic and M31 globular clusters,
and the very few available direct measures of mass-to-light ratios in
early-type galaxies are within these bounds.  The second source of
uncertainty arises from the contribution of the dark matter halo to the
value of $\tau$.  The values of $\kappa_c$ we have used come directly from
models of the lensing configuration, and are relatively insensitive to the
assumed mass profile.  In fact it is seen from Eq.~\ref{eqn1} that
for values of $\kappa_c \lesssim 0.5$, small changes in $\kappa_c$ have
only a small effect on the value of $\tau$.  The third source of error is
from the measurement of surface brightness from the HST frames.  These are
less than 5\% in the value of $\tau$, and small compared with the other
sources of error.

Although there has been widespread interest in simulating magnification
patterns for large optical depths to microlensing where caustic networks
dominate, simulations for low optical depth distributions of lenses have
received less attention.  In order to provide illustrations of the
probability of microlensing at optical depths appropriate to the quasar
images in our sample, we have carried out microlensing simulations to show
the effect of parameters such as $\kappa_c$ and $\gamma$ in low surface
mass distributions.  The sparse distribution of lenses and the minimal
presence of caustic features can be compared with the widespread caustic
network and consequent high probability of microlensing for the optical
depth appropriate to the inner regions of a galaxy.

Taken together, the values of $\tau$ in Table~\ref{tab2} imply that the
probability that most or all of the quasar images are being microlensed by
stars in the lensing galaxy is vanishingly small.  There is perhaps still
some uncertainty over the value of $M/L_H$, in the event that the lensing
galaxies in our sample are not typical of the overall population of giant
ellipticals, although in the case of one of our sample members
HE0435-1223, direct modelling by \cite{w17} gives a value for $M/L_H$ that
is close to our adopted value.  Overall, the errors and uncertainties
discussed above do not significantly change this conclusion.  In
presenting our results we have conservatively excluded RXJ1131-1231 which
has a very low surface mass density of stars in the quasar image
positions, and a very large amplitude of variation due to microlensing.
Inclusion of this system strongly weights our results against the
possibility that the stars in the lensing galaxy can be responsible for
the observed microlensing.  RXJ1131-1231 requires further investigation
to determine whether there is a large component of unseen microlensing
mass associated with the lensing galaxy.

The probability that the quasar images are being microlensed by stars in
the lensing galaxies can also be put on a quantitative footing by the
analysis of magnification maps from microlensing simulations.  This was
achieved by counting the frequency with which randomly generated
difference light curves are microlensed by star fields and smoothly
distributed dark matter components equal to the values for the lensing
galaxies at the positions of the quasar images from Table~\ref{tab2}.
The simulated light curves were assigned a length corresponding to the
time for which each quasar was monitored, and the criterion for being
microlensed was that the amplitude of the microlensing simulation
exceeded that of the observed light curve.  The results listed in
Table~\ref{tab3} confirm the low probability that the amplitudes of the
quasar light curves are the result of microlensing by the stellar
population of the lensing galaxies.  In addition to simulations based on
the values for $\kappa_*$ and $\kappa_c$ from Table~\ref{tab2}, we have
also carried out simulations to show the effect of shear on our results.
It is seen from Table~\ref{tab3} that, as expected, the microlensing
probabilities are mostly increased, but still too low to plausibly account
for the observed microlensing variations.

The effect of source size on microlensing amplitudes has been discussed
in Section~\ref{sec4}, and it is a source of uncertainty in the
microlensing simulations.  The effect of increasing the size of a source
in microlensing simulations is well illustrated by the figures from
\cite{k86}.  This is put on a more quantitative footing by \cite{r91} and
\cite{w94}, who derive analytical expressions to describe the decrease in
amplitude with increasing source size. We have used a disc with half-light
radius of 4 lt-days, which on the basis of recent measurements in the
literature would appear to be the smallest plausible size.  The actual
profile of the source has been shown by \cite{m05b} to be unimportant
compared with the half-light radius in determining the microlensing
amplitude.

Over the last decade or more there have been a number of studies
designed to use microlensing to measure the mass fraction of stars or
compact bodies in galaxy halos.  In Section~\ref{sec6} we have discussed
in some detail two well-known analyses \citep{m09,p12} which conclude that
the observed microlensing variations can best be explained by a relatively
small proportion of compact bodies, consistent with a plausible stellar
component.  This result has been supported by a number of other studies
adopting slightly different approaches (for example \cite{m17}),
but resting on a number of assumptions.  These include the structure of
the dark matter halo, the size of the quasar accretion disc, the mass
function of the microlensing bodies and the extent to which the halo is
clumpy.  In some cases these unknowns are treated as free parameters in a
Bayesian analysis, but more direct measurements have turned out to be very
illuminating.  For example recent studies of stellar dynamics in the Milky
Way \citep{g17,c18} have illustrated the uncertainties on microlensing
constraints from broadening the mass distribution or allowing spatial
clustering for the population of microlenses.  The effect on microlensing
at cosmological distances is still unclear.  However, as we have pointed
out in Section~\ref{sec6}, our values for the mass fraction of stars are
quite similar to these earlier studies.  The main difference between these
studies and our approach is that we have confined our sample to widely
separated systems where the quasar images lie well clear of the bulk of
the stellar distribution.  We then directly estimate the probability that
the observed surface density of stars can be responsible for the
microlensing.  These wide-separation systems tend to give atypical results
in the large samples used for Bayesian analysis.

If stars in the lensing galaxies are not the primary source of the lenses
responsible for the observed microlensing, the question of the true
identity of the lenses must be addressed.  Although a number of
non-particle dark matter candidates have been proposed over the years, our
analysis concludes that the only plausible objects consistent with the
most fundamental constraints are primordial black holes.  In this case the
cosmological distribution of dark matter along the line of sight to the
quasar would also contribute to the optical depth to microlensing.
Whether or not this provides an adequate explanation for the observed
microlensing requires further investigation and more detailed modelling,
but following the general argument of \cite{p73} one can say that there is
a plausible case that dark matter in the form of compact bodies would be
sufficient to account for the observed microlensing in multiply-lensed
quasar systems.  The recent detection of gravitational waves from a black
hole merger \citep{a16} provides further support for the idea that
primordial black holes make up the dark matter \citep{b16,s16,b18},
although there is still ongoing debate as to whether such mergers are
between stellar or primordial black holes.  

\section{Conclusions}

The idea behind this paper is to test the hypothesis that stars in the
lensing galaxy of a multiply lensed quasar system can account
for the observed microlensing of quasar images.  This was achieved by
measuring the surface brightness of starlight at the positions of the
quasar images and converting this into stellar surface mass density.
The surface mass density in smoothly distributed dark matter was estimated
from lensing models of the quasar systems, and applied to the surface mass
density in stars to obtain the optical depth to microlensing $\tau$.
Providing $\tau$ is below a critical value, this is a direct measure of
the probability that a quasar image is microlensed by the stars in the
lensing galaxy.  A sample of four gravitationally lensed quasar systems
were selected for the test, where the quasar images lie on the fringe of
the stellar distribution of the lensing galaxy.  Photometric monitoring
programmes have shown that in each system the images are varying
independently by substantial amounts, implying microlensing by a
population of compact objects.  For all of these systems the optical depth
$\tau$ was found to be below the limiting value where caustic features
form, thus allowing a straightforward estimate of the probability of
microlensing.  The hypothesis that we wish to test is that in at least one
of the systems in our sample the observed microlensing is not caused by
stars in the lensing galaxy, and so by some other population of compact
bodies.  The probability that each system was being microlensed by stars
was calculated to be in the range 2\%-12\%, implying a probability of
around $10^{-6}$ that all four systems were microlensed by stars.
Uncertainties in the assumed mass-to-light ratio and dark halo structure
do not significantly affect this conclusion. There is reason to believe
that RXJ1131-1231 may be untypical in its microlensing properties, and so
if we conservatively exclude it from our results the probability of the
stellar population producing the observed microlensing increases to
$3 \times 10^{-4}$.

Computer simulations of the magnification pattern of lensing galaxy
star fields with the observed light curve tracks superimposed support the
conclusion that the observed microlensing is very unlikely to be produced
by stars. The results from these simulations, again excluding
RXJ1131-1231, imply a probability of $3 \times 10^{-4}$ that the observed
microlensing is due to stars in the lensing galaxy, similar to the result
from the analysis of optical depth to microlensing.  The main uncertainty
in the results from the computer simulations is the assumed size of the
quasar accretion disc, and so we have accordingly adopted a very
conservative value.

We have argued that the most plausible candidates for dark matter in the
form of compact bodies are primordial black holes, and the recent
detection of gravitational waves attributed to a black hole merger
supports this idea.  It is possible that dark matter in the form of
primordial black holes in the lensing galaxy halos are responsible for the
observed microlensing, but it is not clear that they would constitute a
sufficiently large optical depth to microlensing.  A more plausible
possibility is that the microlensing is the result of a cosmological
distribution of dark matter primordial black holes along the line of sight
to the quasar images.

\end{document}